\documentclass[twocolumn,prl,showpacs,nofootinbib,superscriptaddress]{revtex4}
\usepackage{latexsym}
\usepackage{dcolumn}
\usepackage{epsfig}

\def\spai{\sigma_{p{-}\rm air}^{\rm prod}}

\newcommand{\ba}{\begin{eqnarray}}

\newcommand{\ea}{\end{eqnarray}}
\newcommand{\be}{\begin{equation}}
\newcommand{\ee}{\end{equation}}

\newcommand{\eq}[1]{Eq.\,(\ref{#1})}

\newcommand{\pbar}{\bar p}

\newcommand{\sigtot}{\sigma_{\rm tot}}

\def\bea{\begin{eqnarray}} 
\def\eea{\end{eqnarray}}

\begin{document}

\title{New experimental evidence that the proton develops asymptotically into a black disk }

\author{Martin~M.~Block}
\affiliation{Department of Physics and Astronomy, Northwestern University, 
Evanston, IL 60208}
\author{Francis Halzen}
\affiliation{Department of Physics, University of Wisconsin, Madison, WI 53706}

\begin{abstract}
Recently, the  Auger group has extracted the proton-air cross section from observations of air showers produced by cosmic ray protons (and nuclei) interacting in the atmosphere and converted it into measurements of the total and inelastic $pp$ cross sections $\sigma_{\rm tot}$  and $\sigma_{\rm inel}$ at the super-LHC energy of 57 TeV. Their results reinforce our earlier conclusions that the proton becomes a black disk at asymptotic energies, a prediction reached on the basis of sub-LHC $\pbar p$ and $pp$ measurements of $\sigma_{\rm tot}$ and $\rho$, the ratio of the real to the imaginary part of the forward scattering amplitude [M. M. Block and F. Halzen, Phys. Rev. Lett. {\bf 107}, 212002 (2011)]. The same black disk description of the proton anticipated the values of $\sigma_{\rm tot}$  and $\sigma_{\rm inel}$ measured by the TOTEM experiment at the LHC cms (center of mass) energy of $\sqrt s=7$ TeV, as well as those of $\sigma_{\rm inel}$ measured by ALICE, ATLAS and CMS, as well as the ALICE measurement at 2.76 TeV. All data are consistent with a proton that is asymptotically a black disk of gluons: (i) both  $\sigma_{\rm  tot}$  and $\sigma_{\rm inel}$ behave as  $\ln^2s$, saturating the Froissart bound, (ii) the forward scattering amplitude becomes pure imaginary (iii) the ratio $\sigma_{\rm inel}/\sigma_{\rm tot}=0.509 \pm 0.021$, compatible with the black disk value of $\frac{1}{2}$, and (iv) proton interactions become flavor blind.

\end{abstract}
\date{\today}

\maketitle

{\em Introduction.}---Recently, eight measurements of the $pp$ total cross section have been made at energies  beyond the Tevatron. At the LHC cms energy of 2.76 GeV, ALICE \cite{alice}   has measured $\sigma_{\rm inel}$, at the LHC cms  energy of 7 TeV, ALICE \cite{alice}, ATLAS \cite{atlas}, CMS \cite{cms} and TOTEM \cite{totem} have measured $\sigma_{\rm inel}$; $\sigma_{\rm tot}$ has been measured by TOTEM \cite{totem}. Most recently,  the Pierre Auger Observatory has published $pp$ cross sections for $\sigma_{\rm inel}$ and $\sigma_{\rm tot}$ at 57 TeV \cite{auger}. The goal of this note is to update the evidence for the proton asymptotically becoming a black disk of gluons, using these new experimental results.

At 2.76 TeV, the measured value of the $pp$ cross section is:
\begin{description}
\item{ALICE}: $\sigma_{\rm inel}=62.1\pm 1.6({\rm MonteCarlo})\pm 4.3 ({\rm lum.}) $ mb.
\end{description}

At 7 TeV, the measured values are:
\begin{description}
\item{ALICE}: $\sigma_{\rm inel}=72.7\pm 1.1({\rm MonteCarlo})\pm 5.1 ({\rm lum.}) $ mb,
 \item{ATLAS}: $\sigma_{\rm inel}=69.1\pm 2.4({\rm experim.})\pm 6.9 ({\rm extrapol.}) $ mb,
  \item{CMS}:  $\sigma_{\rm inel}=68.0\pm 2.0 ({\rm syst.})\pm 2.4({\rm lum.})\pm 4 ({\rm extrapol.})$ mb,
  \item{TOTEM}: $\sigma_{\rm inel}=73.5\pm 0.6 ({\rm stat.})  \ {}^ {+ 1.8}_{-1.3} ({\rm syst.}) $ mb,
  \item{TOTEM}: $\sigma_{\rm tot}=98.3\pm 0.2 ({\rm stat.})  \pm 2.8 ({\rm syst.}) $ mb.   
\end{description}

At 57 TeV,the measured values are:
\begin{description}
  \item{Auger}: $\sigma_{\rm inel}=92\pm 7 ({\rm stat.})  {}^{+9}_{-11} ({\rm syst.})\pm 7({\rm Glauber}) $ mb, 
\item{Auger}: $\sigma_{\rm tot}=133\pm 13 ({\rm stat.})  {}^{+17}_{-20} ({\rm syst.})\pm 16({\rm Glauber}) $ mb.  
\end{description}

We will show that all high energy measurements are in excellent agreement with recent predictions made by BH (Block and Halzen)  \cite{blackdisk}\, using a combination of an analyticity-constrained \cite{blockanalyticity} fit to sub-LHC data\cite{physicreports,blockhalzen2} to total cross sections and $\rho$-values  in the energy range $6\le\sqrt{s}\le 1800$ GeV, together with an eikonal model\cite{aspen}.

We will first recall the methodology used by BH for obtaining accurate ultra-high energy extrapolations of the total and inelastic $pp$ cross sections. 

{\em Fitting data anchored by analyticity constraints}---Using analyticity constraints  \cite{blockanalyticity} to anchor an analytic amplitude description of $pp$ ($\bar pp$) forward scattering amplitudes \cite{bc},  BH \cite{blackdisk} made accurate predictions of the high energy behavior of both their total cross sections $\sigma_{\rm tot}$ and $\rho$-values,
\ba
\sigma_{\rm tot}&\equiv&{4\pi\over p} {\rm Im}f(\theta_L=0),\\
\rho&\equiv&{{\rm Re}f(\theta_L=0 )\over {\rm Im}f(\theta_L=0)},
\ea
where $f(\theta_L)$ is the $pp$  laboratory scattering amplitude, $\theta_L$, the laboratory scattering angle and $p$ the laboratory momentum. Saturation of the Froissart bound here  means  that the total cross section $\sigtot$ asymptotically behaves as $\ln^2 s$. Furthermore, the use of analyticity constraints allows one to anchor the fits to the accurate  low energy cross section measurements  between 4 and 6 GeV, in the spirit of FESR (Finite Energy Sum Rules) \cite {blockanalyticity}. Making a local fit to the many very accurate experimental values of $\sigma^{\pm}$ between 4 and 6 GeV, for both $\bar pp$ and $pp$, one obtains  \cite{blockhalzen2} fixed ``anchor-points" for $\sigma^{\pm}$ and their energy derivatives in \eq{sigmapmpp} at 6 GeV, the lowest energy value of our analytic amplitude fit. The model parameterizes the even and odd (under crossing) total cross sections and $\rho$-values and fits  4 experimental quantities, $\sigma_{\bar pp}(\nu), \sigma_{p p}(\nu), \rho_{\bar pp}(\nu)$ and $\rho_{p p}(\nu)$ to the high energy analytic amplitude parameterizations \cite{blockhalzen2}
\ba
\sigma^\pm(\nu)&=&\sigma^0(\nu)\pm\  \delta\left({\nu\over m}\right)^{\alpha -1},\label{sigmapmpp}\\
\rho^\pm(\nu)&=&{1\over\sigma^\pm(\nu)}\left\{\frac{\pi}{2}c_1+c_2\pi \ln\left(\frac{\nu}{m}\right)-\beta_{\cal P'}\cot({\pi\mu\over 2})\left(\frac{\nu}{m}\right)^{\mu -1}\nonumber\right.\\
&&\left.+\frac{4\pi}{\nu}f_+(0)\right.
\left.\pm \delta\tan({\pi\alpha\over 2})\left({\nu\over m}\right)^{\alpha -1} \right\}\label{rhopmpp},
\ea
where the upper sign is for $pp$ and the lower sign is for  $\bar pp$,  and, for high energies, ${\nu / m}\simeq{s/ 2m^2}$. The even amplitude cross section is given by
\ba
\sigma^0(\nu)&\equiv&\beta_{\cal P'}\left(\frac{\nu}{m}\right)^{\mu -1}+c_0+c_1\ln\left(\frac{\nu}{m}\right)\nonumber\\
&&+c_2\ln^2\left(\frac{\nu}{m}\right),\label{sig0pp}
\ea 
with $\nu$  the laboratory energy of the incoming proton (anti-proton), $m$ the proton mass, and the `Regge intercept' $\mu=0.5$. Note that $\nu/m=s/2m^2$, so that asymptotically, $\sigma^0\rightarrow \ln^2 s$.

\begin{figure}[h]%Fig. 3
\begin{center}
\mbox{\epsfig{file=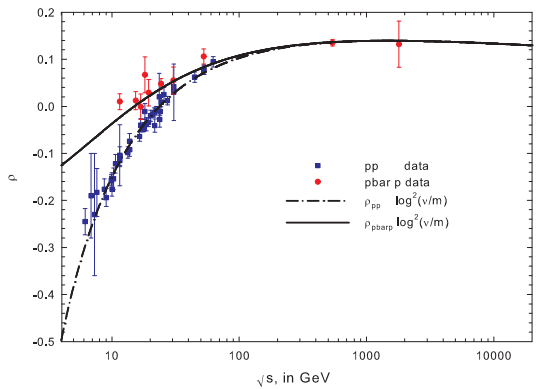
,width=3in%
,bbllx=95pt,bblly=275pt,bburx=535pt,bbury=575pt,clip=%
}}
\end{center}
\caption[ ]{\protect {Froissart-bounded analytic amplitude fits to $\rho$, the ratio of the real to the imaginary portion of the forward scattering amplitude, vs. $\sqrt s$, the cms energy in GeV, taken from BH \cite{blockhalzen2}. 
  The $\bar pp$ data used in the fit  are the (red) circles and the $pp$ data are the (blue) squares. }}
\label{fig:rho}
\end{figure}

The results \cite{blockhalzen2} of a {\em global} fit to both $\bar p p$ and $pp$ $\rho$-values and total cross sections in the energy range $6\le\sqrt s\le 1800$ GeV is shown in Fig. \ref{fig:rho}. 
%%%%%%%%%%%%%%%%%%%%  
As seen from \eq{rhopmpp},  $\rho\rightarrow 0$ as $s\rightarrow\infty$, which is a requirement for a black disk at infinity. However,  the tiny change in $\rho$ from 0.135 at 1800 GeV to 0.132 at 14000 GeV implies that we are nowhere near asymptopia, where $\rho=0$.

\begin{figure}[h,t,b] %Fig.1
\begin{center}
\mbox{\epsfig{file=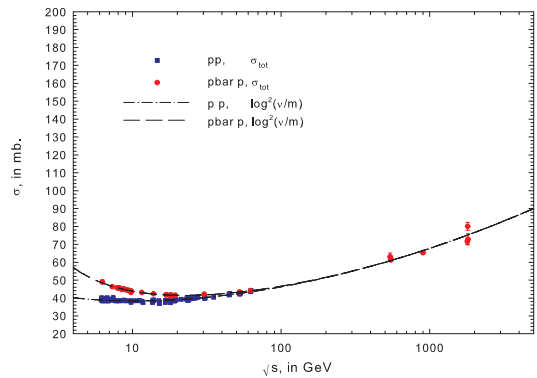
,width=3in%
,bbllx=90pt,bblly=425pt,bburx=555pt,bbury=730pt,clip=%
}}
\end{center}
\caption[]{
 Froissart-bounded analytic amplitude fits to the total cross section, $\sigtot$,  for $\bar pp$ (dashed curve)  and $pp$ (dot-dashed curve)  from \eq{sigmapmpp}, in mb vs. $\sqrt s$, the cms energy in GeV, taken from BH  \cite{blockhalzen2}. The $\bar pp$ data used in the fit  are the (red) circles and the $pp$ data are the (blue) squares.  The fitted data were anchored by values of $\sigtot^{\bar pp}$ and $\sigtot^{pp}$, together with the energy derivatives  ${d\sigtot^{\bar pp}/ d\nu}$ and ${d\sigtot^{pp}/ d\nu}$ at 6 GeV using FESR, as described in Ref. \cite{blockhalzen2}. It should be noted that our ultra-high energy total cross section predictions that are made from our analytic amplitude  fit use {\em only}  total cross section  data that are in the lower energy range  $6\le \sqrt s \le 1800$ GeV.
\label{fig:ppfit}
}
\end{figure}

The fits for the $pp$ and $\bar pp$ total cross sections are shown in Fig. \ref{fig:ppfit}. 
%%%%%%%%%%%%%%%%%%%%%%%%%%%
%%%%%%%%%%%%%%%
The dominant $\ln^2 s$ term in the total cross section $\sigma^0$ (see \eq{sig0pp}) saturates the Froissart bound \cite{froissart}; thus it controls the asymptotic behavior of the cross sections.

Two low energy constraints on $\sigtot^{\bar pp}$ and $\sigtot^{pp}$, together with their energy derivatives  ${d\sigtot^{\bar pp}/ d\nu}$ and ${d\sigtot^{pp}/ d\nu}$ from FESR\cite{blockhalzen2}) can be fixed precisely at 6 GeV by  using the many accurate low energy total cross section measurements between $\sqrt s$ of 4 and 6 GeV. These FESR constraints can be used to fix  the values of  $c_0$ and $\beta_{\cal P'}$, two of the four parameters needed to determine $\sigma^0$, the even high energy total cross section of \eq{sig0pp}.  These values of  $c_0$ and $\beta_{\cal P'}$, together with with the 2 {\em  globally fitted} values of $c_1$ and $c_2$  required for $\sigma^0(\nu)$ (obtained from fitting  the high energy total energy cross section and $\rho$ measurements in the energy region $6\le \sqrt s \le 1800$ GeV), are listed in Table \ref{tab:sigma0}. We remind the reader that only data in the energy region $6\le \sqrt s\le 1800$ GeV are used in this global fit, together with the 4 to 6 GeV total cross section  data used for the  6 GeV low energy `anchor points'. We note that  $c_2$, the coefficient of $\ln^2(s)$, is well-determined, having a statistical accuracy of $\sim 2\%$. Thus, we see from Fig. \ref{fig:ppfit}, that the  experimental data show that a saturated Froissart bound model is accurately satisfied for  total cross sections $\sigtot$ for both $\bar p p$ and for $pp$ in the energy interval $6\le \sqrt s\le 1800$ GeV; this accuracy of prediction mainly results from the use of the FESR constraints on the high energy  analytic amplitude fit \cite{blockanalyticity}. 

Ultra-high energy cross total sections, for which there are no distinction between $\bar p p$ and $pp$ interactions---both being given by $\sigma^0$---are now completely  predicted. For example, we obtain values for the total $pp$ cross section of $\sigma^0=95.4\pm 1.1$ mb at 7 TeV \cite{7tev} and $134.8\pm 1.5$ mb at 57 TeV \cite{cr}, compared to the experimental values of $98.3\pm 2.9$ mb and $133\pm 24$ mb, respectively, where the  systematic and statistical experimental errors have been taken in quadrature.   
%%%%%%%%%%%%%%%%%%%%%%%%%%%%%%%%%%%%%%%%%%%%%%%%
\begin{table}[h,t]                   % Use "table" environment, but also
				 % use  "tabular" environment below.
%
\def\arraystretch{1.5}            % Make the space between rows in the Table,
				  % 1.5 x bigger than the default spacing.

\begin{center}
\caption[]{Values, in mb,  of the parameters needed to determine the even amplitude total cross section, $\sigma^0(\nu)$ of \eq{sig0pp}, taken from Ref. 
\cite{blockhalzen2}; for a fuller description, see the text.
\label{tab:sigma0}
}
\vspace{.2in}
\begin{tabular}[b]{||c||c||}
\hline\hline
$c_0$=$37.32$ mb,&$\beta_{\cal P'}$=$37.10$ mb\\
\hline
$c_1$=$-1.440\pm 0.070$ mb,&$c_2$=$0.2817\pm 0.0064$ mb,\\
\hline\hline
\end{tabular}
     %\vspace{1in} \\
\end{center}
\end{table}
\def\arraystretch{1}  %Restore the default row spacing in the Table.
%%%%%%%%%%%%%%%%%%%%%%%%%%%%%%%%%%%

{\em  Determination of the inelastic cross section}---The inelastic cross section, $\sigma_{\rm inel}^0$, is determined by numerically multiplying the ratio of the inelastic to total cross section with the fitted total cross section $\sigma^0$.  The ratio of inelastic to total cross section was  determined from an eikonal model, called the `Aspen' model; for details see Ref. \cite{aspen} and Ref. \cite{blackdisk}. 

This procedure is purely numerical; when fit by an analytic expression for the  even amplitude high energy inelastic cross section $\sigma^0_{\rm inel}(\nu)$ given by
\ba
\sigma^0_{\rm inel}(\nu)&\equiv&\beta_{\cal P'}^{\rm inel}\left(\frac{\nu}{m}\right)^{\mu -1}+c_0^{\rm inel}+c_1^{\rm inel}\ln\left(\frac{\nu}{m}\right)\nonumber\\
&&+c_2^{\rm inel}\ln^2\left(\frac{\nu}{m}\right),
\ea
we found that 
\ba
\sigma_{\rm inel}^0(\nu)&=& 62.59\left(\frac{\nu}{m}\right)^{-0.5}+24.09+0.1604 \ln\left(\frac{\nu}{m}\right)\nonumber\\
&&+ 0.1433 \ln^2\left(\frac{\nu}{m}\right) \ {\rm mb},
\label{finalinelastic}
\ea 
valid in the energy domain, $\sqrt s \ge 100$ GeV.

\begin{figure}[h]%
\begin{center}
\mbox{\epsfig{file=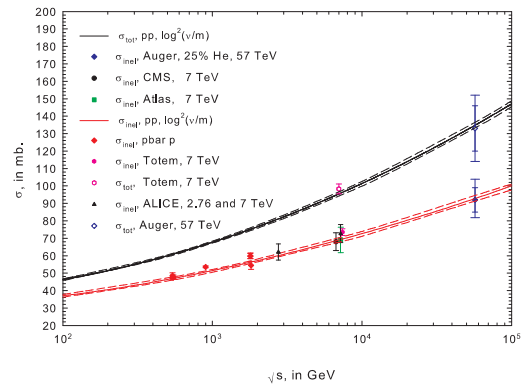
,width=3in%
,bbllx=33pt,bblly=390pt,bburx=510pt,bbury=715
pt,clip=%
}}
\end{center}
\caption[]{\protect
{ Predictions for $\sigma_{\rm tot}$ and $\sigma_{\rm inel}$ vs. $\sqrt s$  for $\bar pp$ and $pp$. For $\sigma_{\rm tot}$, we have compared our predictions with recent $pp$ TOTEM data at 7 TeV and Auger data at 57 TeV, while for $\sigma_{\rm inel}$, we have compared our results with  2.76 TeV $pp$ data from ALICE, 7 TeV $pp$ data from ALICE, ATLAS, CMS and Totem, as well as with the 57 TeV  $pp$ inelastic cross section. The upper solid (black) curve is the central-value prediction for $\sigma_{\rm tot}$ and the lower solid (red) curve is the central-value prediction for $\sigma_{\rm inel}$.  The dotted curves are the errors ($\pm 1 \sigma$) in our predictions, due to the correlated errors of the fitting parameters. We emphasize that {\em none} of the datum points in this plot have been used in our predictions.     
}
}
\label{fig:pppredictions}
\end{figure}

{\em Results}---The lower (red) plot of Fig. \ref{fig:pppredictions} is our prediction  for  high energy  inelastic cross sections $\sigma_{\rm inel}$ as a function of the cms energy in the energy region $100\le \sqrt s\le 100000$ GeV. 
 %%%%%%%%%%%%%%%%%%%%  
%%%%%%%%%%%%%%%%%%%%  
The error bands corresponding to $\pm 1 \sigma$ are the lower dashed curves, and all of the existing inelastic cross section measurements for $\bar p p$, as well as the 5 new ultra-high $pp$ measurements, are shown.  Clearly, the agreement with experiment is excellent over the entire energy scale. 

Also shown as the upper (black) curve in Fig. \ref{fig:pppredictions} is our prediction for ultra-high energy total cross sections $\sigma_{\rm tot}$  (given by $\sigma^0$ of \eq{sig0pp} using the coefficients of Table \ref{tab:sigma0}). Again, the error bands are the upper dashed curves. The excellent agreement with the new highest energy (57 TeV)  experimental measurements is striking. Since {\em none} of the experimental datum points in Fig. \ref{fig:pppredictions} are used in making these predictions, it is clear that our $\ln^2 s$ predictions for $\sigma_{\rm inel}$ and $\sigma_{\rm tot}$ are strongly supported by the existing ultra-high energy measurements. Further confirmation (at 14 TeV) is possible in several years, when the LHC runs at its design energy. 

{\em Asymptopia}---Finally, we \cite{blackdisk} determined the ratio of $\sigma_{\rm inel}(s)/\sigma_{\rm tot}(s)$ as $s\rightarrow \infty$,  given by the ratio of  the  $\ln^2 s$ coefficients  in $\sigma^0_{\rm inel}$ and $\sigma^0$, respectively, i.e.,
\ba  
{\sigma_{\rm inel}\over \sigma_{\rm tot}}\rightarrow {c_2^{\rm inel}\over c_2}={0.1433\over 0.2817}=0.509\pm 0.021,\quad {\rm as\ } s\rightarrow \infty,
\ea
that is well within error of the expected value of $\frac{1}{2}$ that is appropriate for a black disk at infinity.

These new experiments confirm our earlier results \cite{blackdisk} that the asymptotic proton is made up of gluons and thus is flavor blind, yielding  the same asymptotic cross sections for $pp$, $K p$,  $pp$, $\pi p$ and, through vector dominance, to $\gamma p$ and $\gamma^*p$ scattering, as in deep inelastic scattering. We also found \cite{blackdisk} that the coefficient $c_2$  of the $\ln^2 s$ term in the total cross section corresponded to a lowest lying glueball mass of $M_{\rm glueball}=(2\pi/c_2)^{1/2}= 2.97\pm 0.03$ GeV.   

The experimental ratio of $\sigma_{\rm inel}/\sigma_{\rm tot}\approx 0.72$ at 7 TeV; at the highest available energy of 57 TeV, the ratio very slowly decreases to $\approx 0.69$,  not even close to the asymptotic limit of 0.5.  Hence, even at the highest cosmic ray energies,  we still are a very long way from asymptopia and will never get much closer to it. However, it is most interesting that the essential principles of analyticity and unitarity, the underpinnings of  our theoretical results, are experimentally validated up to these ultra-high energies. Clearly, it will become the task of lattice QCD to extend these results to the enormous energies needed to approach asymptopia.  

Additional experimental confirmation that the proton asymptotically approaches a black disk is given by Shegelsky and Ryskin \cite{ryskin}, who analyze all available high energy data on the energy dependence of the  shrinking of the diffraction cone (the nuclear slope parameter $B\equiv d\sigma_{ \rm elastic}/d\ln t|_{t=0}$), finding agreement with our results. 
 
{\em Conclusions}---We find that:
\begin{enumerate}
  \item 
 both the total $pp$ cross section and the inelastic cross section are fit up to $\sqrt s=57$ TeV by a saturated Froissart-bounded $\ln^2s$ behavior that is associated with a black disk.
\item the forward scattering amplitude is pure imaginary as $s\rightarrow \infty$, as is  required for a black disk.
\item the ratio of ${\sigma_{\rm inel}/ \sigma_{\rm tot}}\rightarrow 0.509\pm 0.021,\quad {\rm as\ } s\rightarrow \infty$, compatible with the black disk value of 0.5.
\end{enumerate}
Thus, we conclude that existing experimental evidence strongly supports the conclusion that the proton becomes a black disk at infinity. Our result may have implications for theories of ``new" physics; for instance, string theories whose additional dimensions modify forward scattering amplitudes as shown originally by Amati, Ciafaloni  and Veneziano \cite{amati}.

{\em Acknowledgments}---In part,  F. H. is supported by the National Science Foundation  Grant No. OPP-0236449, by the DOE  grant DE-FG02-95ER40896 and  by the University of Wisconsin Alumni Research Foundation. M. M. B. thanks  the Aspen Center for Physics, supported in part by NSF Grant No. 1066293,  for its hospitality during this work.
%%%%%%%%%%%%%%%%%%%%%%%%%%%%%%%%%%

%
\end{document}